\documentclass{article}

\usepackage{amssymb}
\usepackage[all]{xy} 

\usepackage{theorem}
\newtheorem{theo}{Theorem}[section]
\newtheorem{prop}[theo]{Proposition}

\theorembodyfont{\normalfont}
\newtheorem{defn}[theo]{Definition}
\newtheorem{exam}[theo]{Example}
\newtheorem{rmk}[theo]{Remark}

\newenvironment{resume}
  {\begin{itemize}\item[]\small\textbf{Abstract.}}
  {\normalsize\end{itemize}}

\newcommand{\dv}{v}            
\newcommand{\dc}{c}            
\newcommand{\dt}{c}            
\newcommand{\de}{e}            

\newcommand{\basic}{\mathrm{basic}} 
\newcommand{\deco}{\mathrm{deco}}
\newcommand{\expl}{\mathrm{expl}}
\newcommand{\nat}{\mathrm{nat}}

\newcommand{\Nat}{\mathrm{Nat}} 
\newcommand{\Bool}{\mathrm{Bool}} 
\newcommand{\False}{\mathrm{F}} 
\newcommand{\True}{\mathrm{T}} 
\newcommand{\Unit}{\mathrm{Unit}} 
\newcommand{\Zero}{0}

\newcommand{\ext}{e}                  
\newcommand{\rai}{\mathtt{raise}}           
\newcommand{\han}{\mathtt{handle}}           
\newcommand{\cass}{\mathtt{case}}           
\newcommand{\of}{\mathtt{of}}           
\newcommand{\Exception}{\mathtt{Exception}}           
\newcommand{\id}{\mathtt{id}}     

\newcommand{\bN}{\mathbb{N}}
\newcommand{\bE}{\mathbb{E}}

\newcommand{\suc}{\mathrm{succ}}   
\newcommand{\pre}{\mathrm{pred}}   

\newcommand{\lto}{\longrightarrow}
\newcommand{\To}{\Rightarrow}

\newcommand{\ol}{\overline}
\renewcommand{\circ}{\,.\,}

\title{About raising and handling exceptions}
\author{Dominique Duval and Jean-Claude Reynaud}
\date{April 20., 2006}

\begin{document}

\maketitle

\begin{resume}
This paper presents a unified framework for dealing with 
a deduction system and a denotational semantics of exceptions. 
It is based on the fact that handling exceptions can be seen 
as a kind of generalized case distinction.  
This point of view on exceptions has been introduced in 2004,
it is based on the notion of diagrammatic logic, 
which assumes some familiarity with category theory.
Extensive sums of types 
can be used for dealing with case distinctions.
The aim of this new paper is to focus on the 
role of a generalized extensivity property 
for dealing with exceptions. Moreover, 
the presentation of this paper 
makes only a restricted use of category theory.
\end{resume}

\section{Introduction} 
\label{intro}

This paper presents a unified framework for dealing with a deduction system
and a denotational semantics of exceptions. It is based on the fact that
handling exceptions can be seen as a kind of 
generalized case distinction.
This point of view on exceptions has been introduced in \cite{DR04}, 
and a short presentation can be found in \cite{DR05}. 
In both these papers, some familiarity with category theory 
(adjunction, sketches,\dots) is assumed.
One aim of this new paper to present 
the main ideas of \cite{DR04} in an elementary way,
with a restricted use of category theory.
 
Usual case distinction can be presented 
in a \emph{distributive} logic, 
which means that products and sums of types 
are allowed, and that the 
product is distributive over the sum.
Products and sums of types can be interpreted as 
cartesian products and disjoint unions of sets,
respectively, so that the distributivity property 
does hold on sets.
It follows from \cite{CLW93} that case distinction 
can also be presented in a weaker \emph{extensive} logic,
where sums of types are allowed,  
and the inverse image of a sum by a function is still a sum.
In this paper, exceptions are formalized
in a kind of generalized extensive logic;
in \cite{DR04}, this framework is enriched 
for dealing also with product types.
Exceptions are studied in many different frameworks,
for instance in 
\cite{GDLE83,BBC86,Sc93,BHM02,La02,PP03,WSM05}.
But, to our knowledge, the emphasize on the use of the 
extensivity property for dealing with exceptions, is new.
 
A puzzling issue about exceptions is the apparent discrepancy 
between the deduction system of a language with exceptions 
and its set-valued interpretation.
Indeed, the type of exceptions is implicit in the language,  
while its interpretation requires an explicit set of exceptions.
A major step towards a solution is the use of \emph{monads} in \cite{Mo91},
in the framework of typed lambda-calculus: 
the functions are classified, on the one hand the 
\emph{values} are not allowed to raise any exception,
on the other hand the \emph{computations} may raise 
an exception. So, if the types $X$ and $Y$ are interpreted 
in a set-valued model as the sets $A$ and $B$, 
then a function $f:X\to Y$ is 
interpreted either as a map $\varphi:A\to B$ if $f$ is a value,
or as a map $\varphi:A\to B+\bE$, where $\bE$ is the set 
of exceptional values, if $f$ is a computation.
But this approach fails to formalize in a satisfactory way 
the handling of exceptions in the framework of typed lambda-calculus  
\cite{PP01}.
Our approach succeeds in formalizing the handling of exceptions,
but the extensive logic is fairly different from 
typed lambda-calculus.
Although we do not use monads explicitly, 
we do distinguish values from computations.

Actually, three different extensive logics are presented in
this paper. 
The \emph{basic} extensive logic is described in section~\ref{base}:
there are sums of types, 
and the inverse image of a sum by a function is a sum.
This basic logic does not deal with exceptions.
In the next sections, it is modified in two different ways, 
in order to include a treatment of exceptions.
The \emph{decorated logic with exceptions},
or simply \emph{decorated logic}, is described  
in section~\ref{deco}.
Then the \emph{logic with explicit exceptions},
or simply \emph{explicit logic}, is presented 
in section~\ref{expl}.
Each of both logics for exceptions has its own deduction system 
and denotational semantics,
however the interest of the first one relies primarily in its 
deduction system, 
while the denotational semantics of the second one 
is easier to grasp.
A link between these logics is established,
so that the deduction system of the decorated logic 
is sound with respect to the models in the sense of the explicit logic.
This solves the problem 
of the apparent discrepancy between the deduction system 
of a language with exceptions and its set-valued interpretation.

So, this point of view on exceptions requires 
a framework for dealing with several logics and 
the links between them. 
Such a framework is provided by 
\emph{diagrammatic logics} \cite{DL02,Du03}.
This work does rely on the theory of diagrammatic logics, 
mainly for the definition of the decorated logic
and for the link between the decorated logic and the explicit logic,
as explained in \cite{DR04,DR05}. 
However, in this paper,  
the role of diagrammatic logic is hidden,
and the few required notions about categories are reminded.
Actually, we do not need much more than 
the definition of a category, which is quite simple:
it is a directed graph where the arrows 
can be composed as soon as they are consecutive.
Proofs can be found in \cite{DR05}. 

A diagrammatic logic is well known 
as soon as its \emph{specifications} and \emph{theories} 
are carefully described.
Roughly speaking, a specification is a family of axioms, 
and a theory is a family of theorems that is closed 
under deduction.
The \emph{deduction rules} of the given diagrammatic logic
are used for generating a theory from a specification, 
which means, for deriving theorems from axioms.
The \emph{models} of a specification are then 
defined automatically, in a sound way: 
every theorem that can be proved from a specification 
is satisfied in every model of the specification, 
or equivalently, every model of the specification
can be extended to a model of the generated theory.

\section{About graphs}
\label{graph}

In the three logics that will be described, 
the specifications and theories
are some kind of generalized 
graphs and categories, respectively. 
In this preliminary section, 
we introduce some basic facts about graphs and categories,
that will be used in the next sections.
 
\begin{defn}\textbf{(graph).}
\label{defn:graph-graph}
A \emph{(directed multi-)graph} is made of points 
and arrows, that are called respectively 
\emph{types} $X,Y$,\dots
and \emph{(univariate) functions} $f:X\to Y$,\dots
\end{defn}

A \emph{category} is a graph where functions can be composed,
with the usual properties of composition, as follows.

\begin{defn}\textbf{(category).}
\label{defn:graph-cat}
A \emph{category} is a graph 
where each type has an \emph{identity} function $\id_X:X\to X$, 
each pair of consecutive functions $f:X\to Y$ and $g:Y\to Z$
has a \emph{composed} function $g\circ f:X\to Z$,
and the \emph{unitarity} and \emph{associativity} 
axioms hold (as soon as it makes sense): 
  $$ f\circ\id_X=f \;,\; \id_Y\circ f=f \;,\;
  (h\circ g)\circ f=h\circ (g\circ f) \;.$$
As usual, thanks to associativity, 
parentheses are generally dropped.
\end{defn}

Clearly, each graph generates a category, by adding 
all the missing identities and composed functions, and 
by identifying some functions according to the axioms. 
Generating a category from a graph is similar to 
generating \emph{all} the programs from a grammar of a given language,
or generating \emph{all} the theorems about groups (say)
from a set of axioms for groups.
This is pretty interesting, but far too large:
we are usually quite happy with \emph{some} programs 
and \emph{some} theorems\dots 
More is said about this remark in the 
``decomposition theorem'' of \cite{DL02,Du03}.
About graphs and categories, this remark is the motivation for 
defining something ``between'' both, as follows. 

\begin{defn}\textbf{(compositive graph).}
\label{defn:graph-comp}
A \emph{compositive graph} is a graph 
where each type may have a (potential) 
identity function $\id_X:X\to X$  
and each pair of consecutive functions $f:X\to Y$ and $g:Y\to Z$ 
may have a (potential) composed function $g\circ f:X\to Z$.
\end{defn}

The unitarity and associativity axioms are not mentioned: 
as any equalities, some of them may hold, but this is not 
mandatory.
Typically, a compositive graph may describe a step between 
a graph and its generated category,
when \emph{some} identities and composed functions 
have been generated.

The compositive graphs and the categories form 
the specifications and theories, respectively, 
of a (very simple) diagrammatic logic. 
The rules of this logic are the identity and composition rules,
as well as the rules that correspond to the axioms for categories:
$$ \frac{X}{\id_X:X\to X}\;(\textrm{id}) \qquad 
\frac{f:X\to Y \quad g:Y\to Z}{g\circ f:X\to Z}\;(\textrm{comp}) $$
$$ \frac{f:X\to Y}{f\circ \id_X=f:X\to Y}\;(\textrm{unit}_X) \qquad 
\frac{f:X\to Y}{\id_Y\circ f=f:X\to Y}\;(\textrm{unit}_Y) $$
$$ \frac{f:X\to Y \quad g:Y\to Z\quad h:Z\to T}
{(h\circ g)\circ f=h\circ (g\circ f):X\to T}\;(\textrm{assoc}) $$

The fact that types and functions can be
considered as symbols that stand for sets and maps,
respectively, is catched by the following notion of 
\emph{model}. 
In this paper, only set-valued models are considered; 
a more general definition of models can be found in \cite{DL02,Du03}.
For clarity, we speak about \emph{maps}
(rather than functions) between sets.

\begin{defn}\textbf{(model of a compositive graph).}
\label{defn:graph-mod}
A \emph{(set-valued) model} $M$ of a compositive graph 
interprets each type $X$ as a set $M(X)$
and each function $f:X\to Y$ as a map $M(f):M(X)\to M(Y)$,
in such a way that 
identity functions are interpreted as identity maps 
and composed functions as composed maps:
$M(\id_X)=\id_{M(X)}$ and $M(g\circ f)=M(g)\circ M(f)$.
\end{defn}

\begin{exam}\textbf{(natural numbers).}
\label{exam:graph}
Let us consider the graph  
made of two types $\Unit$ and $\Nat$
and two functions $z:\Unit\to\Nat$ and $s:\Nat\to \Nat$:
  $$ \xymatrix@C=3pc{
  \Unit \ar[r]^{z} & \Nat \ar@(ur,dr)^{s} \\
  }$$
The generated category contains the functions 
$\id_{\Unit}$, $\id_{\Nat}$, as well as 
$s^k:\Nat\to \Nat$ and $s^k\circ z:\Unit\to \Nat$ 
for every $k\in\bN$.  
By adding to the initial graph some of 
these functions, we get a compositive graph. 
The \emph{model of naturals} of all these graphs 
interprets $\Unit$ as a singleton $\{*\}$,  
$\Nat$ as the set $\bN$ of naturals, 
$z$ as the constant map $*\mapsto 0$, 
which is identified to the element $0\in\bN$,
and $s$ as the successor map $\suc:\bN\to\bN$.
Then the function $s^k\circ z$ is interpreted as 
the constant map $*\mapsto k$, identified to $k\in\bN$.
\end{exam}

There is still a technical point to discuss 
about compositive graphs and categories.
Equality between functions 
is often too crude for dealing with computational issues:
for a compiler, functions like $f\circ\id_X$ and $f$ 
are distinct, even though they become identified 
in all models. This is a reason for introducing 
\emph{equations} $f\equiv g$ as \emph{potential equalities}
in compositive graphs:
if $f\equiv g$, then $M(f)=M(g)$ in every model $M$. 
So, from now on, every compositive graph may have equations.

It follows that the categories also have to be modified.
An \emph{equiv-category} looks like a category,
except for two points.
First, it is equipped with equations which form a \emph{congruence},
which means, an equivalence relation compatible with composition.
Second, it satisfies the unitarity and associativity axioms 
only up to congruence. 
For simplicity, and because this will not cause any trouble 
in this paper, we still call it a \emph{category}.

So, this diagrammatic logic is a kind of equational logic, 
where all functions have arity~1.

\section{A basic logic}
\label{base}

In order to focus on the issue of exceptions, 
we have chosen a \emph{basic logic} that 
deals with case distinctions.
As in section~\ref{graph}, all its functions have arity~1, 
since no product of types is provided;  
multivariate functions are considered in \cite{DR04}.
In order to deal with case distinctions, some 
sums of types are needed, 
and they must satisfy a property called \emph{extensivity},
after \cite{CLW93}.
Note that in \cite{CLW93} the word ``extensivity'' 
is used only for categories, while here it is used for sums.
The specifications and theories of the basic logic 
are described below.

\begin{defn}\textbf{(basic specification).}
\label{defn:base-spec}
A \emph{basic specification} $\Sigma$ 
is a compositive graph  
such that some finite lists of types $Y_1,\dots,Y_n$ have a 
\emph{(potential) sum},
made of a \emph{vertex} type $Y_1+\dots+Y_n$ 
and \emph{coprojection} functions 
$j_i:Y_i\to Y_1+\dots+Y_n$, for $i\in\{1,\dots,n\}$.
\end{defn}

\begin{defn}\textbf{(models of a basic specification).}
\label{defn:base-mod}
A \emph{(set-valued) model} of a basic specification 
is a model of the underlying compositive graph
that interprets potential sums as disjoint unions.
\end{defn}

The properties of sums in a basic theory are stated now.
The first one (existence and unicity of matches)
is the usual defining property of sums in a category, 
but only up to congrunce.
The second property (extensivity of sums) will allow to define 
case distinction.

\begin{defn}\textbf{(sums and matches).}
\label{defn:base-sum}
A \emph{sum}  
is a potential sum that satisfies the following property.
If $f_i:Y_i\to Z$, for $i\in\{1,\dots,n\}$, are functions,
then there is a \emph{match} $[j_1\To f_1\mid\dots\mid j_n\To f_n]$ 
or $[f_1\mid\dots\mid f_n]:Y_1+\dots+ Y_n\to Z$,
i.e., a function such that
$[f_1\mid\dots\mid f_n]\circ j_i\equiv f_i$ 
for $i\in\{1,\dots,n\}$,
and if $f:Y_1+\dots+ Y_n\to Z$ is a function 
such that $f\circ j_i\equiv f_i$ for $i\in\{1,\dots,n\}$ 
then $f\equiv [f_1\mid\dots\mid f_n]$.
\end{defn}
The existence of matches can be illustrated as follows, when $n=2$,
with dotted arrows for representing the coprojections:
  $$  
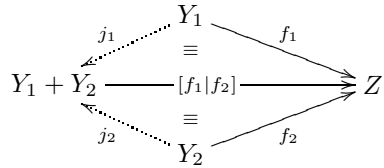
\xymatrix@R=1pc@C=2pc{ 
    & Y_1 \ar@{..>}[dl]_{j_1} \ar[drr]^{f_1} \ar@{}[d]|{\equiv} && \\ 
    \quad Y_1+Y_2 \ar[rrr]|{\,[f_1\mid f_2]\,} & \mbox{} & \mbox{} & Z \\ 
    & Y_2 \ar@{..>}[ul]^{j_2} \ar[urr]_{f_2} \ar@{}[u]|{\equiv} && \\ 
  } $$

When $n=0$, a sum ``of no type'' is 
called an \emph{initial} type, denoted $\Zero$; 
it satisfies the following property.
If $Z$ is a type,
then there is a function $[\;]_Z:\Zero\to Z$ such that,
if $f:\Zero\to Z$ is a function,
then $f\equiv [\;]_Z$.
The existence of empty matches can be illustrated as follows:
  $$\xymatrix@R=1pc@C=2pc{ 
    \Zero \ar[rr]|{\,[\;]_Z\,} && Z \\ 
  }$$

\begin{defn}\textbf{(the inverse image of a sum by a function).}
\label{defn:base-inv}
Let $Y=Y_1+\dots+ Y_n$ be a sum, with coprojections $j_1,\dots,j_n$, 
and let $u:X\to Y$ be a function.
An \emph{inverse image of the sum $Y=Y_1+\dots+ Y_n$
by the function $u$} is a sum $X=u^{-1}(Y_1)+\dots+u^{-1}(Y_n)$,
with coprojections $u^{-1}(j_1),\dots,u^{-1}(j_n)$,
together with \emph{restriction} functions 
$u_i:u^{-1}(Y_i)\to Y_i$
such that, for $i\in\{1,\dots,n\}$:
  $$ j_i\circ u_i\equiv u\circ u^{-1}(j_i) \;.$$
\end{defn}
Here is an illustration when $n=2$.
  $$ 
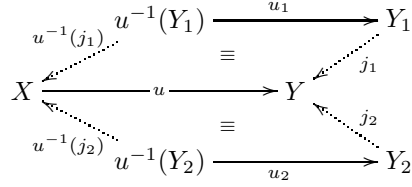
\xymatrix@R=1pc@C=2pc{ 
    & u^{-1}(Y_1) \ar@{..>}[dl]_{u^{-1}(j_1)} \ar[rr]^{u_1} \ar@{}[dr]|{\equiv}
    && Y_1 \ar@{..>}[dl]^(.4){j_1} \\ 
    X \ar[rr]|{\,u\,}  
    && Y & \\ 
    & u^{-1}(Y_2) \ar@{..>}[ul]^{u^{-1}(j_2)} \ar[rr]_{u_2} \ar@{}[ur]|{\equiv} 
    && Y_2 \ar@{..>}[ul]_(.4){j_2} \\ 
  } $$

\begin{defn}\textbf{(extensivity).}
\label{defn:base-exte}
A sum $Y=Y_1+\dots+Y_n$ is \emph{extensive} 
if, for every function 
$u:X\to Y$ there is an inverse image of the sum $Y=Y_1+\dots+ Y_n$
by the function $u$, and it is unique
(the unicity of inverse images, here and in the sequel,
is only up to some equivalence).
\end{defn}

\begin{defn}\textbf{(basic theories).}
\label{defn:base-theory}
A \emph{basic theory} $\Theta$ 
is a basic specification
such that its underlying graph is a category, 
and all its potential sums of types are extensive sums.
\end{defn}

The category of sets can be seen as a basic theory, 
with the equality for congruence.
It is not assumed here that all sums of types do exist in a 
basic theory, although this property could be added.
Now, case distinction in any basic theory 
is easily defined, thanks to the properties of sums. 

\begin{defn}\textbf{(cases).}
\label{defn:base-case}
Let $Y=Y_1+\dots+ Y_n$ be a sum, 
$u:X\to Y$ a function, 
and let $X=u^{-1}(Y_1)+\dots+u^{-1}(Y_n)$ be the inverse image. 
Let $f_i:u^{-1}(Y_i)\to Z$ be functions, for $i\in\{1,\dots,n\}$.
The \emph{case distinction} function 
(or simply the \emph{case} function) that acts as 
$f_i$ on $u^{-1}(Y_i)$, for all $i$, is:
$$ \cass\;u\;\of\;[\, j_i\To f_i\,]_{1\leq i\leq n}
 \;=\; [\,u^{-1}( j_i)\To f_i\,]_{1\leq i\leq n}\;:\; X \to Z \;. $$
\end{defn}
  $$  \xymatrix@R=1pc@C=2pc{ 
    & u^{-1}(Y_1) \ar@{..>}[dl]_{u^{-1}(j_1)} 
       \ar[drr]^{f_1} \ar@{}[d]|{\equiv} && \\ 
    \quad X \ar[rrr]|{\,\cass\,u\,\of[j_1\To f_1\mid j_2\To f_2]\,} & 
      \mbox{} & \mbox{} & Z \\ 
    & u^{-1}(Y_2) \ar@{..>}[ul]^{u^{-1}(j_2)} 
        \ar[urr]_{f_2} \ar@{}[u]|{\equiv} && \\ 
  } $$
This means that the case function is characterized by the equations:
  $$ (\cass\;u\;\of\;[\, j_i\To f_i\,]_{1\leq i\leq n}) \circ
      (u^{-1}( j_i)) \equiv f_i \;\mbox{ , for }\; 1\leq i\leq n \;.$$
Clearly, when $u=\id_Y:Y\to Y$, then the case function is 
congruent to a match:
  $$(\cass\;\id_Y\;\of\;[\, j_i\To f_i\,]_{1\leq i\leq n})
   \equiv [\, j_i\To f_i\,]_{1\leq i\leq n} \;:\; Y \to Z \;.$$

The basic specifications and the basic theories form 
a diagrammatic logic, in this paper it is called 
the \emph{basic} logic. 
The rules of this logic are the identity and composition rules,
as in section~\ref{graph}, 
together with the rules for the 
existence and unicity of matches
and for the extensivity of sums.

\begin{rmk}\textbf{(booleans).}
In order to recover a type of booleans, 
a sum $\Bool=\False+\True$ can be used.
Then a function with values in $\Bool$ is called 
a \emph{predicate}.
The inverse image of the sum 
$\Bool=\False+\True$ by a predicate $p:X\to\Bool$
is also a sum, say $X=X_b+X_{\ol{b}}$, 
because of the extensivity property.
In the basic theory of sets,
it can be assumed that the types $\False$ and $\True$ are 
interpreted as singletons, so that $\Bool$ is 
interpreted as the usual set of booleans.
Then, in every model $M$, 
the sets $M(X_b)$ and $M(X_{\ol{b}})$ are 
the parts of $M(X)$ where the map $M(b)$ is true and false, 
respectively.
\end{rmk}

\begin{exam}\textbf{(the basic specification $\Sigma_{\nat}$).}
\label{exam:base}
The graph in example~\ref{exam:graph} can be 
considered as a basic specification, 
with no equation and no sum.
The rules of the basic logic can be used for
deriving, for instance, 
the functions $[s\To s\circ s\mid z\To z]:\Nat\to\Nat$,
and (the subscript $\Nat$ is omitted):
  $$ p = \cass\;\id\;\of\;[\,s\To \id \mid z\To z\,] \equiv
  [\,s\To \id \mid z\To z\,] :\Nat\to \Nat \;.$$
From its definition, the function $p$ satisfies the 
equations $p\circ z \equiv z$ and $p\circ s\equiv \id$.
As in example~\ref{exam:graph},
we are interested in the model of naturals 
of $\Sigma_{\nat}$, called $M_{\nat}$.
In this model, the function $p$ must be interpreted as the 
predecessor map $\pre:\bN\to\bN$ such that $\pre(n)=n-1$
for each positive $n$ and $\pre(0)=0$.
\end{exam}

\section{A decorated logic for exceptions}
\label{deco}

\subsection{Three keywords for exceptions}
\label{deco-key}

We use the keywords $\rai$ for raising exceptions 
and $\han$ for handling them, as in Standard ML.  

The predecessor map $\pre:\bN\to\bN$ from example~\ref{exam:base} 
can also be formalised in the following way, 
if some mechanism for exceptions is available:
\begin{itemize}
\item[] First, an exception $\ext$ is created: 
  $$ \Exception\;\ext $$
\item[] Then, a function $p':N\to N$ is generated, 
  such that $p'(z)$ raises the exception $\ext$:
  $$ p'(x) = \cass\;x\;\of\;[\,s(y)\To y \mid z \To \rai \;\ext\,] $$
\item[] Finally, a function $p'':N\to N$ is generated, 
  that calls $p'$ and handles the exception $\ext$:
  $$ p''(x) = p'(x)\;\han\;[\,\ext\To z\,] $$
\end{itemize}

The basic logic is now modified, in order to be able to deal with
the mechanism of exceptions, with its three keywords:
  $$ \Exception \,,\quad \rai \,,\quad \han \;.$$
For this purpose, we use a kind of logic where the functions 
are \emph{decorated}: each function is 
associated to a symbol, which is called its \emph{decoration}, 
and which appears as a superscript.
The decorations are ``$\dv$'' for \emph{value} 
and ``$\dc$'' for \emph{computation},
they are borrowed from the monads approach \cite{Mo91}.
What is new here, is that the rules of the logic 
are also decorated, as will be explained below.
In particular, various decorations of the extensivity property
will give rise to various kinds of case distinctions, 
which in turn will be used for formalizing the treatment of exceptions.
We claim that expressions of the form:
  $$ \rai\;\;\ext \qquad\mbox{ or }\qquad f\;\;\han\;\;g \;,$$
can be considered as decorated functions;
the keywords ``$\rai$'' and ``$\han$'' are constructors 
for new decorated functions, 
very much like ``$[\dots]$'' and ``$\cass$'' 
are constructors for new basic functions.
Moreover, the decoration of every function can be easily 
derived from the use of the keyword ``$\Exception$'' 
and from the rules of the decorated logic, as follows:
every exception is a computation, 
and every function involving a computation is a computation.

One issue with the decorated logic is that it does not 
have set-valued models in such a simple way as the 
basic logic in section~\ref{base}
or the explicit logic in section~\ref{expl},
which blurs the intuition about this logic. 
In section~\ref{expl}, the decorated logic will be mapped 
to the explicit logic, and a set-valued interpretation 
will then be recovered. 

\begin{exam}\textbf{(the decorated specification $\Sigma_{\nat,\deco}$).}
\label{exam:deco-key}
In the next examples, a decorated specification $\Sigma_{\nat,\deco}$ 
is built progressively, so that
a predecessor decorated function $p$ is defined 
in example~\ref{exam:deco-dv} without using exceptions, 
then a predecessor decorated function $p''$ is defined 
in example~\ref{exam:deco-dv} with the help of exceptions,   
and finally (also in example~\ref{exam:deco-dv})
it is proved, in the decorated logic, that $p''$ is congruent to~$p$.
\end{exam}

\subsection{The decoration ``$\dv$'' for ``value''}
\label{deco-dv}

The functions that 
have nothing to do with the exceptions are called \emph{values};
they are decorated with the symbol~$\dv$, i.e., 
the notation $f^{\dv}$ means that the function $f$ is a value.
An equation between values is called a \emph{value equation}, i.e.,
the notation $f\equiv^{\dv} g$ means that 
$f^{\dv}\equiv g^{\dv}$ is an equation between values.
The identities are values, and the composition of values is a value.
The value equations generate a congruence.
The sums of types behave as in the basic logic,
with values instead of arbitrary functions:
the coprojections are values,  
a match of values is a value,
and the extensivity property holds for values,
so that cases over values give rise to values.
These sums, matches and cases are denoted 
as in the basic logic, in particular
the initial type for values is denoted $\Zero$.
For the case construction, this means that 
a case like ``$\cass\;u\;\of\;[\, j_i\To f_i\,]_i$'', 
where $u$ and the $f_i$'s are values, is the value:
  $$ (\cass\;u^{\dv}\;\of\;
     [\, j_i^{\dv}\To f_i^{\dv}\,]_{1\leq i\leq n})^{\dv} 
    \;=\; (\cass\;u\;\of\;
     [\, j_i\To f_i\,]_{1\leq i\leq n})^{\dv} 
    \;=\; 
    { [\,u^{-1}( j_i)\To f_i\,]_{1\leq i\leq n} }^{\dv}  \;. $$
So, one rule of the decorated logic is the extensivity rule for values,
which says that every sum has a unique inverse image by every value. 
For binary sums, this rule can be illustrated as follows.
  $$ \xymatrix@R=1pc@C=2pc{ 
    & u^{-1}(Y_1) \ar@{..>}[dl]_{(u^{-1}(j_1))^{\dv}} \ar[rr]^{u_1^{\dv}} \ar@{}[dr]|{\equiv}
    && Y_1 \ar@{..>}[dl]^(.4){j_1^{\dv}} \\ 
    X \ar[rr]|{\,u^{\dv}\,}  
    && Y & \\ 
    & u^{-1}(Y_2) \ar@{..>}[ul]^{(u^{-1}(j_2))^{\dv}} \ar[rr]_{u_2^{\dv}} \ar@{}[ur]|{\equiv} 
    && Y_2 \ar@{..>}[ul]_(.4){j_2^{\dv}} \\ 
  } $$

\begin{exam}\textbf{(the value part of $\Sigma_{\nat,\deco}$).}
\label{exam:deco-dv}
In our example, the value part of the decorated specification 
$\Sigma_{\nat,\deco}$
is a copy of the basic specification $\Sigma_{\nat}$ from
example~\ref{exam:base}. 
Hence, $\Sigma_{\nat,\deco}$ has two types $\Unit$ and $\Nat$, 
two values $z^{\dv}:\Unit\to\Nat$ and $s^{\dv}:\Nat\to\Nat$, 
and no value equation.
It generates a value:
  $$ p^{\dv} = (\cass\;\id\;\of\;[\,s\To \id \mid z\To z\,])^{\dv}
    \equiv [\,s\To \id \mid z\To z\,]^{\dv} : \Nat\to\Nat $$
so that $p\circ s \equiv^{\dv} \id$ and $p\circ z \equiv^{\dv} z$.
\end{exam}

\subsection{The decoration ``$\dc$'' for ``computation''}
\label{deco-dc}

All the functions that may raise exceptions
are called \emph{computations}; 
they are decorated with the symbol~$\dc$, 
as well as the equations between them.
Since computations \emph{may} (and not \emph{must}) 
raise exceptions, 
each value $f^{\dv}$ may be \emph{coerced} into a computation $f^{\dc}$,
and similarly each value equation may be coerced into a computation equation.
The composition of computations yields a computation,
and the computation equations generate a congruence.
In the composed computation $(g\circ f)^{\dc}=g^{\dc}\circ f^{\dc}$,
it is expected that any exception which is raised by $f^{\dc}$ 
is propagated by $g^{\dc}$: this is proved in theorem~\ref{theo:deco-raise}.

A match of computations is a computation, 
in a straightforward way.
When $n=0$, this means that 
the initial type for values is also initial for computations:
for every type $X$, 
there is a unique value $[\,]_X^{\dv}:\Zero\to X$, 
and its coercion as a computation $[\,]_X^{\dc}$ is 
the unique computation $[\,]_X^{\dc}:\Zero\to X$.

Since a match of computations is a computation, 
a case like ``$\cass\;u\;\of\;[\, j_i\To f_i\,]_i$'' 
is defined when the $f_i$'s are computations and $u$ is a value;
the same notation ``$\cass$'' is used for this construction:
  $$ (\cass\;u^{\dv}\;\of\;
     [\, j_i^{\dv}\To f_i^{\dc}\,]_{1\leq i\leq n})^{\dc} 
    \;=\;  (\cass\;u\;\of\;
     [\, j_i\To f_i\,]_{1\leq i\leq n})^{\dc} 
    \;=\; {[\,u^{-1}( j_i)\To f_i\,]_{1\leq i\leq n}}^{\dc} \;. $$
But there is no such definition when $u$ is a computation;
indeed, if $u$ raises an exception, 
there is no canonical way to decide which $Y_i$ the exception ``comes from''.
However, in section~\ref{deco-casec} a special situation is described, 
where some kind of ``$\cass\;u^{\dc}\;\of\;\dots$'' 
can be defined, when $u$ is a computation.

\subsection{The keyword $\Exception$}
\label{deco-exc}

In a decorated specification, the values are generated from 
some elementary values, which are the operation symbols of a signature,
and the computations are generated from 
some elementary computations, which are the \emph{exceptions}.
Recall that a computation $f^{\dc}:X\to Y$ in a decorated specification
may raise an exception instead of returning a result of type $Y$. 
Following this idea, we consider that 
a declaration ``$\Exception\;\ext\;\of\;P$'', for any type $P$, 
adds to the decorated specification a computation 
$\ext^{\dc}:P \to \Zero$:
indeed, such a computation cannot return a result 
of type $\Zero$, since $\Zero$ stands for the empty set, 
hence it has to raise an exception.
  $$ \xymatrix@C=4pc{
    P \ar[r]^{\ext^{\dc}} & \Zero \\
  } $$ 
In this paper, for simplicity,
it is assumed that all the exceptions in a 
decorated specification are given once and for all.
The exceptions form the coprojections 
of a new kind of sum in the decorated specification;
this \emph{exceptional sum} is studied in section~\ref{deco-casee}.

\begin{exam}\textbf{(the exception of $\Sigma_{\nat,\deco}$).}
\label{exam:deco-exc}
In the decorated specification $\Sigma_{\nat,\deco}$,
the declaration ``$\Exception\;\ext$'' adds a computation 
$\ext^{\dc}:\Unit \to \Zero $, 
from which other computations will be derived 
in example~\ref{exam:deco-raise}.
\end{exam}

\subsection{The keyword $\rai$}
\label{deco-raise}

Recall that $\Zero$ is an initial type for values and for computations.
We claim that when a function $f:X\to Y$
\emph{raises} an exception $\ext$, this means that 
the exception $\ext$ can be viewed as an expression of type $Y$. 
This is expressed in the following definition.
 
\begin{defn}\textbf{(the keyword $\rai$).}
\label{defn:deco-raise}
The keyword $\rai$ is the polymorphic value:
  $$ {\rai_Y}^{\dv} = {[\,]_Y}^{\dv} : \Zero \lto Y \;.$$
In a decorated specification $\Sigma$, 
let $\ext^{\dc}:P\to\Zero$ be an exception and $Y$ a type.
To \emph{raise the exception $\ext^{\dc}$ in the type $Y$}
is to build the composition:
  $$ (\rai_Y \circ \ext)^{\dc} : P \lto Y \;.$$
\end{defn}

The following result proves that the exceptions propagate, as required;
it is a consequence of the unicity of the empty sum.

\begin{theo}\textbf{(propagation of exceptions).}
\label{theo:deco-raise}
For every computations $f^{\dc}:X\to \Zero$ and $g^{\dc}:Y\to Z$
(typically, when $f^{\dc}$ is an exception): 
  $$ g\circ \rai_Y \circ f \equiv^{\dc} \rai_Z \circ f \;.$$
\end{theo}

\begin{exam}\textbf{(raising an exception in $\Sigma_{\nat,\deco}$).}
\label{exam:deco-raise}
In the decorated specification $\Sigma_{\nat,\deco}$,
the computation $p'$ is defined as follows:
  $$ {p'}^{\dc} = 
  (\,\cass\;\id\;\of\;[\,s\To \id \mid z\To \rai\circ\ext\,]\,)^{\dc}  
    \equiv [\,s\To \id \mid z\To \rai\circ\ext\,]^{\dc} : \Nat\to\Nat $$
It follows from theorem~\ref{theo:deco-raise} that, 
for every computation $g^{\dc}:\Nat\to\Nat$, 
the computation $(g\circ p'\circ z)^{\dc}$ raises the exception $\ext$.
\end{exam}

\subsection{The case construction over a computation}
\label{deco-casec}

The case construction over a computation, which is described now, 
can be used only inside 
a handle construction (section~\ref{deco-handle}).
Such a construction occurs only with respect to a sum 
of the form $Y+\Zero$, for any type $Y$.
It is easy to prove that the vertex of this sum is isomorphic to $Y$,  
with the coprojections $\rai_Y$ and $\id_Y$
(the subscript $Y$ is often omitted):
indeed, the proof involves only values, it is similar to the usual proof
in the basic logic.
This sum $Y=Y+\Zero$ may be used as the other sums, 
for building 
matches of values and matches of computations, 
and also for building inverse images of values,
but this has little interest: 
the inverse image of the sum $Y=Y+\Zero$ by a value $u^{\dv}:X\to Y$ 
is simply the sum $X=X+\Zero$.
The interesting property of the sum $Y=Y+\Zero$
is that there is a special rule for it:
this sum has an inverse image by every 
\emph{computation} $u^{\dc}:X\to Y$.
Indeed, if $u$ raises an exception, then we decide that this exception 
``comes from'' the $\Zero$ part of the sum $Y=Y+\Zero$.
More precisely, this inverse image is defined below.

\begin{defn}\textbf{(the inverse image of a sum by a computation).}
\label{defn:deco-invc}
Let $u^{\dc}:X\to Y$ be a computation.
An \emph{inverse image of the sum $Y=Y+\Zero$
by the computation $u$} is a sum $X=X_{u,1}+X_{u,0}$,
with value coprojections $j_{u,1}^{\dv}:X_{u,1}\to X$
and $j_{u,0}^{\dv}:X_{u,0}\to X$,
and with a \emph{value} $u_1^{\dv}:X_{u,1}\to Y$ and
a \emph{computation} $u_0^{\dc}:X_{u,0}\to\Zero$
such that:
  $$ u\circ  j_{u,1} \equiv^{\dc} u_1 \;\mbox{ and }\,
     u\circ  j_{u,0} \equiv^{\dc} \rai_Y\circ u_0 \;. $$
\end{defn}
  $$\xymatrix@R=1pc@C=2pc{ 
    & X_{u,1} \ar@{..>}[dl]_{j_{u,1}^{\dv}} \ar[rr]^{u_1^{\dv}} 
       \ar@{}[dr]|{\equiv^{\dc}} 
    && Y \ar@{..>}[dl]^(.4){\id^{\dv}}  \\
    X \ar[rr]|{\,u^{\dc}\,}  
    && Y & \\ 
    & X_{u,0} \ar@{..>}[ul]^{j_{u,0}^{\dv}} \ar[rr]_{u_0^{\dc}} 
       \ar@{}[ur]|{\equiv^{\dc}}
    && \Zero \ar@{..>}[ul]_(.4){\rai^{\dv}} \\
  } $$

Some properties of this inverse image are stated now,
their proof is easy. 
The second one shows that there is no ambiguity in our definition:
when a computation $u^{\dc}$ comes, by coercion, from a value $u^{\dv}$,
then the inverse image of $Y=Y+\Zero$ by the computation $u$
is the same as the inverse image of $Y=Y+\Zero$ by the value $u$.
The last property proves the ``back-propagation'' 
of the raising of exceptions, with respect to values:
if $u'\circ u$ raises an exception,
and if $u$ is a value, then $u'$ raises the same exception.

\begin{prop}\textbf{(properties of the inverse image 
of a sum by a computation).}
\label{prop:deco-invc}
\mbox{}
\begin{itemize}
\item Let $u^{\dc},{u'}^{\dc}:X\to Y$ be two computations 
such that $u\equiv^{\dc}u'$, then
  $u^{-1}(Y+\Zero) = {u'}^{-1}(Y+\Zero) \;.$
\item Let $u^{\dv}:X\to Y$ be a value, then
  $ (u^{\dv})^{-1}(Y+\Zero) = X+\Zero \;.$
\item Let $u^{\dc}:X\to Y$ be a computation 
  and ${u'}^{\dv}:Y\to Z$ a value, then
  $ ({u'}^{\dv}\circ u)^{-1}(Z+\Zero)= u^{-1}(Y+\Zero) \;.$
\item Let $u^{\dc}:X\to Y$ be a computation such that 
  $u\equiv^{\dc}\rai_Y\circ f$ for some computation 
  $f^{\dc}$, then $(u^{\dv})^{-1}(Y+\Zero) = \Zero+X \;.$
\item Let $u^{\dc}:X\to Y$ be a computation and ${u'}^{\dv}:Y\to Z$ a value,
  such that $u'\circ u \equiv^{\dc} \rai_Z\circ f$ for some computation 
  $f^{\dc}$, then $u \equiv^{\dc} \rai_Y\circ f$.
\end{itemize}
\end{prop}

\begin{defn}\textbf{(extensivity for computations).}
\label{defn:deco-extec}
A sum $Y=Y+\Zero$ is \emph{extensive for computations} 
if, for every computation 
$u^{\dc}:X\to Y$ there is an inverse image of the sum $Y=Y+\Zero$
by the computation $u$, and it is unique.
\end{defn}

The rule of extensivity for computations
states that in a decorated theory, for every type $Y$
the sum $Y=Y+\Zero$ is extensive for computations.

\begin{defn}\textbf{(cases over computations).}
\label{defn:deco-casec}
Let $u^{\dc}:X\to Y$ be a computation, 
and $f_1^{\dc}:X_{u,1}\to Z$ and $f_0^{\dc}:X_{u,0}\to Z$ two computations.
Then the computation 
``$\cass^{\dt}\;u\;\of\;[\,\id\To f_1\mid\rai\To f_0\,]$'', 
which is called a \emph{case over computation} construction, is defined as: 
$$ (\,\cass^{\dt}\;u^{\dc}\;\of\;[\,\id\To f_1\mid\rai\To f_0\,]\,)^{\dc}
 \;=\; [\, j_{u,1}\To f_1 \mid  j_{u,0}\To f_0\,]^{\dc} \;:\; X \to Z \;. $$
\end{defn}
This means that the case over computation function 
is characterized by the equations:
  $$ (\cass^{\dt}\;u\;\of\;[\,\id\To f_1\mid\rai\To f_0\,] \circ
      (u^{-1}( j_i)) \equiv f_i \;\mbox{ , for }\; 1\leq i\leq n \;.$$

\subsection{The exceptional case construction}
\label{deco-casee}

Let us come back to the declarations of exceptions.
The exception declarations 
``$\Exception\;\ext_i\;\of\;P_i$'', for $1\leq i\leq k$, 
add to the decorated specification a sum of a new kind,
called the \emph{exceptional sum}, 
which allows to test which one among the $\ext_i$'s is some 
given exception.
From now on, let:
  $$ \ext_i^{\dc}:P_i\to\Zero \mbox{, for } 1\leq i\leq k\;,$$
be the exceptions in some given decorated specification.

\begin{defn}\textbf{(the exceptional sum).}
\label{defn:deco-sume}
The \emph{exceptional sum} $\Zero=\sum_{i=1}^kP_i$ has vertex $\Zero$ 
and coprojections the computations $\ext_i^{\dc}$'s for $1\leq i\leq k$.
\end{defn}

The exceptional sum is quite special: 
its coprojections are computations, instead of values,
and it is used only inside a handle construction (section~\ref{deco-handle}).
The exceptional sum enjoys a decorated version of only one 
among the properties of sums, namely the extensivity, as follows.

\begin{defn}\textbf{(the inverse image of the exceptional sum
by a computation).}
\label{defn:deco-inve}
Let $u^{\dc}:X\to\Zero$ be a computation. 
An \emph{inverse image of the exceptional sum by $u^{\dc}$} 
is a sum $X=\sum_{i=1}^ku^{-1}(P_i)$, 
with \emph{values} coprojections $(u^{-1}(\ext_i)^{\dv}$, 
together with \emph{values} $u_i^{\dv}:u^{-1}(P_i)\to P_i$
such that for each $i$:
  $$ u\circ (u^{-1}(\ext_i)) \equiv^{\dc} \ext_i \circ u_i \;. $$
\end{defn}
  $$\xymatrix@R=1pc@C=2pc{ 
    & u^{-1}(P_1) \ar@{..>}[dl]_{u^{-1}(\ext_1)^{\dv}} 
       \ar[rr]^{u_1^{\dv}} \ar@{}[dr]|{\equiv^{\dc}}
    && P_1 \ar@{..>}[dl]^(.4){\ext_1^{\dc}} \\ 
    X \ar[rr]|{\,u^{\dc}\,} && \Zero & \\ 
    & u^{-1}(P_2) \ar@{..>}[ul]^{u^{-1}(\ext_2)^{\dv}} 
       \ar[rr]^{u_2^{\dv}} \ar@{}[ur]|{\equiv^{\dc}}
    && P_2 \ar@{..>}[ul]_(.4){\ext_2^{\dc}} \\ 
  } $$

\begin{defn}\textbf{(extensivity of the exceptional sum).}
\label{defn:deco-extee}
The exceptional sum is \emph{extensive} 
if it has a unique inverse image 
by every computation with type $\Zero$.
\end{defn}

The rule of extensivity for exceptions 
states that in a decorated theory, 
the exceptional sum is extensive.

Now the exceptional case construction can be defined,
as another decorated version of the basic case construction.

\begin{defn}\textbf{(exceptional cases).}
\label{defn:deco-casee}
Let $u^{\dc}:X\to\Zero$ be a computation,
$I$ a subset of $\{1,\dots,k\}$,
and for each $i\in I$ let $f_i^{\dc}$ be a computation:
  $$ f_i^{\dc} : u^{-1}(P_i)\to Y \;.$$ 
For each $i\not\in I$, let $f_i^{\dc}$ be the \emph{default} computation:
  $$ f_i^{\dc} = (\rai_Y \circ u \circ u^{-1}(\ext_i) )^{\dc} 
    :u^{-1}(P_i)\to Y \;.$$
Then the computation 
``$\cass^{\de}\;u\;\of\;[\,\ext_i\To f_i\,]_{i\in I} $'', 
which is called an \emph{exceptional} case construction, is defined as: 
$$ (\,\cass^{\de}\;u^{\dc}\;\of\;[\,\ext_i\To f_i\,]_{i\in I}\,)^{\dc} 
   \;=\; {[\,u^{-1}(\ext_i)\To f_i  \,]_{1\leq i\leq k}}^{\dc}
   \;:\; X \to Y \;. $$
\end{defn}
This means that the computation 
``$\cass^{\de}\;u\;\of\;[\,\ext_i\To f_i\,]_{i\in I} $''
is characterized by the equations:
  $$ \cass^{\de}\;u\;\of\;[\,\ext_i\To f_i\,]_{i\in I}\circ 
      (u^{-1}(\ext_i)) \equiv f_i \;\mbox{ , for }\; 1\leq i\leq k \;.$$

\begin{exam}\textbf{(an exceptional case in $\Sigma_{\nat,\deco}$).}
\label{exam:deco-casee}
In the decorated specification $\Sigma_{\nat,\deco}$,  
there is only one exception $\ext_1=\ext$,
so that $k=1$ and $P_1=\Unit$, in the exceptional sum.
We may consider the computations $u^{\dc}=\ext^{\dc}:\Unit\to\Zero$
and: 
$$ w^{\dc} =  \cass^{\de}\;u\;\of\;[\,\ext\To z\,] : \Unit\to\Nat \; .$$ 
Then clearly $u^{-1}(\ext)=\id_{\Unit}$, so that
$ w^{\dc} \equiv^{\dc} z : \Unit\to\Nat$.
\end{exam}

\subsection{The keyword $\han$}
\label{deco-handle}

The keyword ``$\han$'' has two arguments: 
for instance, in the function ``$p'\;\han\;[\,\ext\To z\,]$'', 
the arguments of $\han$ are $p'$ and $[\,\ext\To z\,]$.
There are two nested kinds of cases in a handling expression
``$f\;\han\;g$''.
The first one tests whether $f$ raises an exception,
and when this is true, the second one tests which is the raised exception.
The first one is a case distinction over a computation, 
as in section~\ref{deco-casec},
and the second one is an exceptional case distinction, 
as in section~\ref{deco-casee}.
Now, the handling construction is easily defined 
from these two kinds of cases. 

\begin{defn}\textbf{(the keyword $\han$).}
\label{defn:deco-handle}
Let $u^{\dc}:X\to Y$ be a computation,
and let $X=X_{u,1}+X_{u,0}$ 
be the inverse image of the sum $Y=Y+\Zero$ by the computation $u^{\dc}$,  
together with the restrictions 
$u_1^{\dv}:X_{u,1}\to Y$ and $u_0^{\dc}:X_{u,0}\to \Zero$.
Let $X_{u,0}=\sum_{i=1}^k u_0^{-1}(P_i)$ 
be the inverse image of the exceptional sum 
by the computation $u_0^{\dc}$.
Let $I$ be a subset of $\{1,\dots,k\}$ and 
for each $i$ in $I$, let $f_i^{\dc}:u_0^{-1}(P_i)\to Y$ be a computation.
To \emph{handle an exception arising from $u^{\dc}$ according to the match 
$[\,\ext_i\To f_i\,]_{i\in I}$} is to build the computation:
  $$ (u\;\han\;[\,\ext_i\To f_i\,]_{i\in I})^{\dc}
     = (\cass^{\dt}\;u\;\of\;[\,\id_Y \To u_1 \mid \rai_Y \To f\,])^{\dc} 
     : X\lto Y \;, $$
where $f$ is the computation:
  $$ f^{\dc} 
     = (\cass^{\de}\;u_0\;\of\;[\,\ext_i\To f_i\,]_{i\in I})^{\dc} 
    : X_{u,0}\lto Y \;. $$
\end{defn}

The following result proves that the exceptions are handled as required;
it can be compared to the rules for ``$\han$'' in the definition of SML
\cite{MTH90}.

\begin{theo}\textbf{(properties of the handling of exceptions).}
\label{theo:deco-handle}
\mbox{}
\begin{itemize}
\item Let $u_1\equiv^{\dc}u_2:X\to Y$, 
then (with the above notations):
  $$ u_1\;\han\;[\ext_i\To f_i]_{i\in I} \equiv^{\dc} 
     u_2\;\han\;[\ext_i\To f_i]_{i\in I} \;.$$
\item For every value $u^{\dv}:X\to Y$:
  $$ u\;\han\;[\ext_i\To f_i]_{i\in I} \equiv^{\dc} u \;.$$
\item For every computation $u^{\dc}=\rai_Y\circ u':X\to Y$ 
where ${u'}^{\dc}:X\to \Zero$:
  $$ u\;\han\;[\ext_i\To f_i]_{i\in I} \equiv^{\dc}  
     \cass^{\de}\;u'\;\of\;[\,\ext_i\To f_i\,]_{i\in I} \;. $$
If in addition $u'=\ext_j\circ u'':X\to Y$ for some $j\in\{1,\dots,k\}$  
and some value ${u''}^{\dv}:X\to P$, then:
  $$ u\;\han\;[\ext_i\To f_i]_{i\in I} \equiv^{\dc} f_j \;\mbox{ if }\; j\in I \;,$$
  $$ u\;\han\;[\ext_i\To f_i]_{i\in I} \equiv^{\dc} u \;\mbox{ otherwise }\;.$$
\end{itemize}
\end{theo}

\begin{exam}\textbf{(handling an exception in $\Sigma_{\nat,\deco}$).}
\label{exam:deco-handle}
From example~\ref{exam:deco-dv}, $p^{\dv}$ is the value:
  $$ p^{\dv} = \cass\;\id\;\of\;[\,s\To \id \mid z\To z\,] 
     \equiv [\,s\To \id \mid z\To z\,] : \Nat\to\Nat \;.$$
On the other hand, from example~\ref{exam:deco-raise}, 
${p'}^{\dc}$ is the computation:
  $$ {p'}^{\dc} = \cass\;\id\;\of\;[\,s\To \id \mid z\To \rai\circ\ext\,] 
     \equiv [\,s\To \id \mid z\To \rai\circ\ext\,] : \Nat\to\Nat \;. $$
Now, let:
  $$ {p''}^{\dc} = p'\;\han\;[\,\ext\To z\,] : \Nat\to\Nat \;,  $$ 
As an example of a proof in the decorated logic, let us prove that
$ p'' \equiv^{\dc} p$.

It follows from the definition of ${p'}^{\dc}$ that:
  $$ \xymatrix@R=1pc@C=2pc{ 
    & \Nat \ar@{..>}[dl]_{s^{\dv}} \ar[rr]^{\id^{\dv}} 
       \ar@{}[dr]|{\equiv^{\dc}}   
    && \Nat \ar@{..>}[dl]^(.4){\id^{\dv}} \\ 
    \Nat \ar[rr]|(.6){\,{p'}^{\dc}\,}  
    && \Nat & \\ 
    & \Unit \ar@{..>}[ul]^{z^{\dv}} \ar[rr]_{\ext^{\dc}} 
       \ar@{}[ur]|{\equiv^{\dc}}
    && \Zero \ar@{..>}[ul]_(.4){\rai^{\dv}} \\ 
  } $$
Hence,
the inverse image of the sum $\Nat=\Nat+\Zero$ by the computation $p'$
is $\Nat=\Nat+\Unit$, with coprojections $s$ and $z$,
and with ${p'_1}^{\dv}=\id$ and ${p'_0}^{\dc}=\ext$.
Thus:
  $$ {p''}^{\dc} = p'\;\han\;[\,\ext\To z\,]
      \equiv^{\dc} \cass^{\dt}\;p'\;\of\;[\,\id \To \id \mid \rai \To w \,] 
    \equiv^{\dc} [\,s \To \id \mid z \To w\,] : \Nat\to\Nat \;,  $$ 
where, as in example~\ref{exam:deco-casee}:
  $$ w^{\dc} = \cass^{\de}\;\ext\;\of\;[\,\ext\To z\,] 
     \equiv^{\dc} z : \Unit\to\Nat \; .$$ 
It follows that:
  $$ p'' \equiv^{\dc} [\,s \To \id \mid z \To z\,] : \Nat\to\Nat \;.$$
Finally, from the unicity of matches, we conclude that:
  $$ p'' \equiv^{\dc} p \;.$$
Since $p$ is a value, it follows that the computation $p''$, 
actually, never raises an exception.
\end{exam}

\subsection{Undecoration}
\label{deco-dec}

\begin{defn}\textbf{(undecoration).}
The \emph{undecoration} of a decorated specification $\Sigma_{\deco}$ 
is the basic specification $\Sigma_{\basic}$
that is obtained simply by forgetting the decorations. 
\end{defn}

In the framework of diagrammatic logics,
it is easy to check that the undecoration is a morphism
from the decorated logic to the basic logic.

By undecoration, every value $f^{\dv}:X\to Y$ or 
computation $f^{\dc}:X\to Y$ in $\Sigma_{\deco}$
gives rise to a function $f:X\to Y$ in $\Sigma_{\basic}$.
Decorated sums and cases in $\Sigma_{\deco}$, 
give rise to ordinary sums and cases in $\Sigma_{\basic}$.
The sum $X=X_{u,1}+X_{u,0}$ gives rise to the sum $X=X+\Zero$,
and the exceptional sum to a sum with vertex~$\Zero$.
Hence, the undecoration allows to get a simplified
view on the functions and equations, by forgetting all the decorations.
It allows to get a simplified view on the proofs,
since the image of a proof in the decorated logic is a proof 
in the basic logic. This can be stated as: 
\\ $\null\qquad$ \textsl{``A proof in $\Sigma_{\deco}$ is  
a proof in $\Sigma_{\basic}$ which can be decorated''.}
\\ This yields a two-step method for checking a proof 
in the decorated logic: first, the proof without its decorations 
must be valid in the basic logic, then it must be feasible 
to add the decorations in a way that is valid in the decorated logic.

However, this simplified view ``does not preserve the meaning'':
for instance, when $\Unit$ is interpreted as a singleton,
a constant exception $\ext^{\dc}:\Unit\to\Zero$ 
in $\Sigma_{\deco}$ gives rise in $\Sigma_{\basic}$ 
to a function $\ext:\Unit\to\Zero$, 
which has no set-valued interpretation.
In section~\ref{expl}, the \emph{expansion}
of a decorated specification is defined; 
it is more subtle than the undecoration,
and it ``does preserve the meaning''.

\begin{exam}\textbf{(the undecoration of $\Sigma_{\nat,\deco}$).}
\label{exam:deco-dec}
By undecorating $\Sigma_{\nat,\deco}$, 
we get a basic specification $\Sigma_{\nat,\basic}$, 
with a function $\ext:\Unit\to\Zero$,
so that this basic specification has no set-valued model
where $\Unit$ is interpreted as a singleton.
The computation ${p''}^{\dc}$ in $\Sigma_{\nat,\deco}$, 
that involves the three kinds of decorated cases,
gives rise in $\Sigma_{\nat,\basic}$ to a function  
that involves three times the basic case distinction.
\end{exam}

\section{A logic with explicit exceptions}
\label{expl}

\subsection{Expansion}
\label{expl-exp}

The exceptions are now considered in an \emph{explicit} way,
which means that there is a type of exceptions~$E$ which 
formalizes the set of exceptions,
and that $E$ appears in the type of a function, 
as soon as this function may raise an exception.
This corresponds to the \emph{explicit} logic, which has no decorations.
It is an enrichment of the basic logic with a distinguished type $E$.

\begin{defn}\textbf{(explicit specification).}
An explicit specification is a basic specification 
together with a distinguished type $E$.
\end{defn}

\begin{defn}\textbf{(expansion).}
The \emph{expansion} of a decorated specification $\Sigma_{\deco}$
is the explicit specification $\Sigma_{\expl}$ 
obtained by adding the distinguished type $E$,
keeping each value $f^{\dv}:X\to Y$ as a function $f:X\to Y$,  
and replacing each computation $f^{\dc}:X\to Y$ 
by a function $f:X\to Y+E$. 
\end{defn}

In the framework of diagrammatic logics,
it is easy to check that the expansion is a morphism
from the decorated logic to the explicit logic.

So,  
every non-exceptional sum $\sum_{i=1}^n ( j_i^{\dv}:Y_i\to Y)$ 
in $\Sigma_{\deco}$ 
gets expanded as a sum $\sum_{i=1}^n ( j_i:Y_i\to Y)$ in $\Sigma_{\expl}$.
The initial type $\Zero$ in $\Sigma_{\deco}$ 
gets expanded as the initial type $\Zero$ in $\Sigma_{\expl}$,
and the value $\rai_Y^{\dv}=[\;]^{\dv}:\Zero\to Y$
gets expanded as $[\;]:\Zero\to Y$, for each type $Y$. 
In this way, the properties of sums of values in the decorated logic
get satisfied by their images in the explicit logic.
This includes the existence and unicity  
of the inverse image of any value $u^{\dv}$, which gets expanded 
as the inverse image of the function $u$. 
This also includes the property that there are matches of computations; 
indeed let $(f_i^{\dc}:Y_i\to Z)_{1\leq i\leq n}$ 
be computations in $\Sigma_{\deco}$, 
they get expanded as functions $(f_i:Y_i\to Z+E)_{1\leq i\leq n}$,
and the computation $[f_1\mid\dots\mid f_n]^{\dc}:Y \to Z$
gets expanded as the function $[f_1\mid\dots\mid f_n]:Y \to Z+E$.

For the cases over computations,  
let $u^{\dc}:X\to Y$ be a computation in $\Sigma_{\deco}$,
then the expansion of the inverse image of the sum 
$Y=Y+\Zero$ by the computation $u^{\dc}$ 
is the inverse image of the sum 
$Y+E$ by the function $u:X\to Y+E$ in $\Sigma_{\expl}$.

For the exceptional cases,
the exceptions $\ext_i^{\dc}:P_i\to\Zero$ 
get expanded as $\ext_i:P_i\to E$. 
So, the expansion of the exceptional sum 
is the sum $E=\sum_{i=1}^kP_i$, with coprojections the $\ext_i$'s,
and the expansion of an inverse image of the exceptional sum 
is an inverse image of this sum.

Since the raising and handling of exceptions have been defined 
in terms of these decorated case constructions, 
they get expanded accordingly.

\begin{exam}\textbf{(the expansion of $\Sigma_{\nat,\deco}$).}
\label{exam:expl-exp}
Let $\Sigma_{\nat,\expl}$ be the expansion of $\Sigma_{\nat,\deco}$:
it is made of a copy of $\Sigma_{\nat}$ from
example~\ref{exam:base}, together with $\ext:\Unit\to E$,
which has to be a sum, which means that $\ext$ has to be invertible.
\end{exam}

\subsection{Models}
\label{expl-deno}

Let $\Sigma_{\deco}$ be a decorated specification,
and $\Sigma_{\expl}$ the explicit specification 
obtained by expanding $\Sigma_{\deco}$.
Let $\bE$ be a fixed set, called the \emph{set of exceptions}. 
A \emph{(set-valued) model of $\Sigma_{\expl}$ with set of exceptions $\bE$}
is defined as a (set-valued) model (in the basic sense)
such that the interpretation of the distinguished type $E$ is the set $\bE$.
So, the exceptions $\ext_i^{\dc}:P_i\to \Zero$ in $\Sigma_{\deco}$,
that are expanded as $\ext_i:P_i\to E$ in $\Sigma_{\expl}$,
are interpreted as maps $M(\ext_i):M(P_i)\to\bE$.
It follows that $\bE$ must be the disjoint union of the $M(P_i)$'s.

It follows, as required, that the models of the expanded 
specifications provide a denotational semantics for 
the decorated logic. 

\begin{theo}\textbf{(soundness).}
\label{theo:expl-deno}
The deduction system of the decorated logic is sound 
with respect to the explicit denotational semantics.
\end{theo}

This means that every equation of $\Sigma_{\deco}$
(either between values or between computations) 
is interpreted as an equality in every model 
of $\Sigma_{\expl}$. 
A proof of this result can be found in \cite{DR04},
it relies upon the fact that the decorated and the explicit
logics can be formalized as diagrammatic logics, 
and that the expansion is a morphism between them.

\begin{exam}\textbf{(the expansion of $\Sigma_{\nat,\deco}$).}
\label{exam:expl-deno}
Let $\bE=\{\varepsilon\}$.
Then $\Sigma_{\nat,\expl}$ has a model $M_{\nat,\expl}$ 
that interprets $\Unit$, $\Nat$, $z$, and $s$ 
as $\{*\}$, $\bN$, 0 and $\suc$, respectively, 
and $\ext:\Unit\to E$ as $\varepsilon:\{*\}\to\bE$. 
In this model, the computation $(\rai_{\Nat}\circ\ext)^{\dc}$ 
and the value $z^{\dv}$
are interpreted respectively as $\varepsilon$ and $0$.
The value $p^{\dv}$ is interpreted as 
the predecessor map $\pre:\bN\to\bN$,
such that $\pre(n)=n-1$ for $n>0$ and $\pre(0)=0$.
The computation ${p'}^{\dc}$ is interpreted as 
the map $\pre':\bN\to\bN+\bE$,
such that $\pre'(n)=n-1$ for $n>0$ and $\pre'(0)=\varepsilon$.
And the computation ${p''}^{\dc}$ is interpreted as 
the map $\pre'':\bN\to\bN+\bE$,
such that (like $\pre$) 
$\pre''(n)=n-1$ for $n>0$ and $\pre''(0)=0$.
\end{exam}

\section{Conclusion}

Two logics for dealing with exceptions are presented
in this paper. 
The decorated logic provides a deduction system,
and the explicit logic provides a denotational semantics.
The expansion, from the decorated logic to the explicit logic, 
ensures soundness.

Perspectives include the comparison of this approach 
with other formalizations. 
Another direction for future research is to use a similar 
approach, via morphisms of diagrammatic logics, in order to 
study other computational effects; 
in particular, the combination of various effects 
should run smoothly in our diagrammatic framework.

\end{document}